








\magnification 1200
\overfullrule 0 pt




\font\abs=cmr9
\font\ist=cmr8

\font\uit=cmu10


\def\CcC{{\hbox{\tenrm C\kern-.45em{\vrule height.67em width0.08em depth-.04em
\hskip.45em }}}}
\def\RrR{{\hbox{\tenrm I\kern-.17em{R}}}}
\def\HhH{{\hbox{\tenrm {I\kern-.18em{H}}\kern-.18em{I}}}}
\def\DdD{{\hbox{\tenrm {I\kern-.18em{D}}\kern-.36em {\vrule height.62em
width0.08em depth-.04em\hskip.36em}}}}
\def\IiI{{\hbox{\tenrm I\kern-.19em{I}}}}
\def\NnN{{\hbox{\tenrm {I\kern-.18em{N}}\kern-.18em{I}}}}
\def\QqQ{{\hbox{\tenrm {{Q\kern-.54em{\vrule height.61em width0.05em
depth-.04em}\hskip.54em}\kern-.34em{\vrule height.59em width0.05em
depth-.04em}}
\hskip.34em}}}
\def\OoO{{\hbox{\tenrm {{O\kern-.54em{\vrule height.61em width0.05em
depth-.04em}\hskip.54em}\kern-.34em{\vrule height.59em width0.05em
depth-.04em}}
\hskip.34em}}}

\def\uq2{U_q({\uit su}(2))}

\def\fraz#1#2{{\strut\displaystyle #1\over\displaystyle #2}}

\def\part#1{\fraz{\partial}{\partial#1}}

\def\ii#1{\item{$\phantom{1}#1~$}}

\def\su2q{SU(2)_q}
\def\h1q{H(1)_q}

\def\nu{N_{1}}

\hsize= 15 truecm
\vsize= 22 truecm
\hoffset= 0.5 truecm
\voffset= 0 truecm

\null\vskip1.5truecm

\baselineskip= 13.75 pt
\footline={\hss\tenrm\folio\hss}

\centerline
{\bf  QUANTUM DISSIPATION AND QUANTUM GROUPS}
\smallskip
\bigskip
\centerline{
{\it
Alfredo Iorio and Giuseppe Vitiello }}
\bigskip

{\ist Dipartimento di Fisica - Universit\`a di Salerno and INFN--Napoli,
I84100 Salerno, Italy}
\footnote{}{\hskip -.85truecm {\abs E--mail: vitiello@sa.infn.it
{}~~~~~~~~~~~~~~~Annals of Physics (N.Y.), in press.
\hfill}}

\bigskip
\bigskip
\bigskip
\bigskip
\bigskip
\bigskip
\bigskip

\noindent{\bf Abstract.} {\abs We
discuss the r\^ole of  quantum deformation of
Weyl-Heisenberg algebra in dissipative
systems and finite temperature systems. We express the time evolution
generator of the damped harmonic oscillator and the generator of thermal
Bogolubov transformations in terms of operators of the quantum Weyl-Heisenberg
algebra.
The quantum parameter acts as a label for the unitarily inequivalent
representations of the canonical commutation relations in which
the space of the states splits in the infinite volume limit.}

\smallskip
\noindent

PACS 02.20.+b; 03.65.Fd; 11.30.Qc;

\vfill\eject

\noindent {\bf 1. Introduction}
\bigskip

The attention which has been devoted to quantum
deformations$^{[1,2]}$ of
Lie algebras in the last years is motivated
by their appearance in  many problems
of great physical interest, in statistical mechanics as well as in
nonlinear dynamical systems, in conformal theories as well as
in solid state physics. Recently, it has been also recognized$^{[3]}$ that
quantum deformations appear whenever a discrete (space or time) length
characterizes the system under study.

Our aim  in the present paper is to explicitly express the
time evolution generator of the damped harmonic oscillator
and the generator of the thermal Bogolubov
transformations in terms of operators of the
quantum deformation of the Weyl-Heisenberg
algebra. We thus
establish the relation of quantum Weyl-Heisenberg
algebra with
the hamiltonian of the damped harmonic oscillator and with
thermal field theories.

One aspect of physical interest which emerges from
our study relies in the fact that the
quantum deformation parameter turns out to be related with time
parameter in the
case of damped harmonic oscillator and  with temperature in the case of
thermal field theory. In both cases, the quantum parameter acts
as a label for the unitarily inequivalent representations
of the canonical commutation relations
in which the space of the states splits
in the infinite volume limit
(in the case of damped harmonic oscillator  the unitarily inequivalent
representations are
spanned by the time evolution operator, in the case of finite
temperature field theory they are spanned by the generator of
the Bogolubov transformations). Such a conclusion confirms
a general analysis
$^{[4]}$ which shows that the Weyl representations
in Quantum Mechanics  and the
unitarily inequivalent  representations in Quantum Field
Theory are indeed labelled by the deformation parameter of
the quantum Weyl-Heisenberg algebra. In some sense the present results,
providing explicit physical examples, confirm and clarify the r\^ole
of quantum deformations
in specific
applications as dissipative systems and finite temperature systems.

In refs. 5-8 some aspects of dissipation in Quantum Field
Theory
have been studied by considering the canonical quantization of damped
harmonic oscillator
and it has been proven that
the space of the physical states splits into unitarily inequivalent
representations of the canonical commutation relations
in the infinite volume limit.
Also, it has been realized that canonical quantization of the
damped harmonic oscillator
leads to $SU(1,1)$ time-dependent coherent states,
which are well known
in high energy physics as well as in quantum optics
and thermal field theories. Moreover, dissipation phenomena
and squeezed$^{[9]}$ coherent states have been mathematically
related, thus showing their
common physical features$^{[5]}$. In Sec. 2 we will summarize
the formalism for the canonical quantization of the damped harmonic oscillator.

Quantum deformations of Lie algebras are well studied mathematical
structures and therefore their properties need not to be presented
and discussed again in this paper; we only recall that they are
deformations of the enveloping algebras of Lie algebras and have
Hopf algebra structure. In particular the quantum Weyl-Heisenberg
algebra has the
properties of graded Hopf algebra$^{[10]}$. In Sec. 3 we will present
the realization of quantum Weyl-Heisenberg algebra in terms of finite
difference operators over the set of the entire analytic functions
$^{[3]}$.
In this work we do not study the r\^ole of the coproduct operation
neither we investigate the superalgebra
features of quantum Weyl-Heisenberg algebra in connection with our physical
examples. We plan to study
such topics in the future.

Finally, in Sec. 4, the
formal relation between the damped harmonic oscillator
Hamiltonian and the quantum Weyl-Heisenberg
algebra is established. The relation of quantum Weyl-Heisenberg
algebra with thermal field theory at real time (thermo field dynamics
$^{[11,12]}$) is also shown in Sec. 4.
Sec. 5 is devoted to the conclusions.
A preliminary discussion on the subject of the present paper has
been presented in ref. 13.

\bigskip
\bigskip
\noindent {\bf 2. The canonical quantization of the damped
harmonic oscillator}
\bigskip

\indent The Feshbach
and Tikochinsky $^{[14]}$ quantum hamiltonian is
$$
H = \hbar\Omega(\alpha^{\dagger}\beta + \alpha \beta^{\dagger})
- {i\hbar\gamma\over{4m}}\left[(\alpha^{2} - \alpha^{\dagger 2})
 - (\beta^{2} - \beta^{\dagger 2})\right]  ~~,
\eqno{(2.1)}
$$
where
$$
\eqalign{
\alpha &\equiv \left ({1\over{2 \hbar \Omega}} \right )^{1\over{2}} \left (
{{p_{x}}\over{\sqrt{m}}} - i \sqrt{m} \Omega x \right ) \quad , \quad
\alpha^{\dagger} \equiv \left ({1\over{2 \hbar \Omega}} \right )^{1\over{2}}
\left ( {{p_{x}}\over{\sqrt{m}}} + i \sqrt{m} \Omega x \right ) \quad , \cr
\beta &\equiv \left ({1\over{2 \hbar \Omega}} \right )^{1\over{2}} \left (
{{p_{y}}\over{\sqrt{m}}} - i \sqrt{m} \Omega y \right ) \quad , \quad
\beta^{\dagger} \equiv \left ({1\over{2 \hbar \Omega}} \right )^{1\over{2}}
\left ( {{p_{y}}\over{\sqrt{m}}} + i \sqrt{m} \Omega y \right ) \quad ,}
\eqno{(2.2)}
$$

$$
[\, \alpha , \alpha^{\dagger} \,] = 1 = [\, \beta , \beta^{\dagger} \, ] \quad
, \quad  [\, \alpha , \beta \,] = 0 = [\, \alpha , \beta^{\dagger} \, ] \quad .
\eqno{(2.3)}
$$
Of course,
$[\, x , p_{x} \, ] = i\, \hbar = [\, y , p_{y} \,]~ , \quad
[\, x , y \,] = 0 = [\, p_{x} , p_{y} \, ]$ ;  $\Omega $ is the frequency.
The modes $\alpha$ and $\beta$
refer to the damped oscillator with classical motion equation:
$$
m \ddot x + \gamma \dot x + \kappa x = 0 \quad , \eqno{(2.4)}
$$
and to its {\it time-reversed} image:
$$
m \ddot y - \gamma \dot y + \kappa y = 0 \quad ,  \eqno{(2.5)}
$$
respectively.
As discussed in ref. 14 and refs. 5-8,
the introduction of the $\beta$ (or $y$) oscillator is required in order
to set up the canonical quantization scheme.
In ref. 8 it has been shown that, at quantum level, the $\beta$ modes
may allow quantum noise effects arising from the imaginary part of the action.
By using the canonical linear transformations
$
\displaystyle{A \equiv {1\over{\sqrt 2}}
( \alpha + \beta )}
$
and
$
\displaystyle{B \equiv {1\over{\sqrt 2}}
( \alpha - \beta )}~,
$
$H$ may also be written as
$$
\eqalign{
 H &=  H_{0} +  H_{I} \quad , \cr
 H_{0} = \hbar \Omega ( A^{\dagger} A - B^{\dagger} B ) \quad &, \quad
H_{I} = i \hbar \Gamma ( A^{\dagger} B^{\dagger} - A B ) \quad ,\cr}
\eqno{(2.6)}
$$
\noindent where the decay constant for the classical variable $x(t)$ is
denoted by $\Gamma \equiv {{\gamma}\over{2 m}}$.

We note that the states generated by $B^{\dagger}$ represent
the sink where the energy
dissipated by the quantum damped oscillator flows: the $B$-oscillator
represents the (single-mode) reservoir or heat bath coupled to the
$A$-oscillator.

The dynamical group structure associated
with the system of coupled quantum oscillators is that of $SU(1,1)$.
We also observe that
$
[\, H_{0} , H_{I}\, ] = 0
$.

The time-evolution of the vacuum
$
|0> \equiv | n_{A} = 0 , n_{B} = 0 > ~,~
A |0> = 0 = B |0>~
$,
is controlled by $H_{I}$
$$
\eqalign{
|0(t)> = \exp{ \left ( - i t {H \over{\hbar}} \right )} |0> =
\exp{ \left ( - i t {H_{I} \over{\hbar}} \right )} |0>   \cr
= {1\over{\cosh{(\Gamma t)}}} \exp{
\left ( \tanh {(\Gamma t)} A^{\dagger} B^{\dagger} \right )}|0> \quad ,}
\eqno{(2.7)}
$$
$$
<0(t) | 0(t)> = 1~ \quad \forall t~,
\eqno{(2.8)}
$$
$$
\lim_{t\to \infty} <0(t) | 0> \, \propto \lim_{t\to \infty}
\exp{( - t  \Gamma )} = 0 \quad .
\eqno{(2.9)}
$$
Eq. (2.9) express the instability (decay) of the vacuum under the evolution
operator $\exp{ \left ( - i t {H_{I} \over{\hbar}} \right )}$.
We note that  $| 0(t)>$ is a two-mode Glauber coherent state$^{[15,16]}$.

Time evolution transformations for creation and annihilation operators are
$$
\alpha \mapsto \alpha(t) =
{\rm e}^{- i {t\over{\hbar}} H_{I}}
\alpha ~{\rm e}^{i {t\over{\hbar}} H_{I}} =
\alpha \cosh{(\Gamma t)} - \alpha^{\dagger} \sinh{(\Gamma t)} ~,
 \eqno{(2.10a)}
$$
$$
\beta \mapsto \beta(t) =
{\rm e}^{- i {t\over{\hbar}} H_{I}}
\beta ~{\rm e}^{i {t\over{\hbar}} H_{I}} =
\beta \cosh{(\Gamma t)} + \beta^{\dagger} \sinh{(\Gamma t)}
\quad   \eqno{(2.10b)}
$$
and h.c., and the corresponding ones for $A(t)$, $B(t)$ and h.c..
We note that (2.10) are canonical transformations preserving the
canonical commutation relations (2.3).

\indent The signs of $\sinh{(\Gamma t)}$ in (2.10a) and (2.10b)
show that $\alpha$ and $\beta$ are each other time-reversed.
{}From eq. (2.1) we see that
the time-evolution operator ${\cal U}(t)$ may be written as
$$
{\cal U}(t) = \exp{ \left ( - i t {H_{I} \over{\hbar}} \right )} =
\exp\Bigl(-{\Gamma t \over{ 2}}\bigl({\alpha}^2 -
{\alpha}^{\dagger 2}\bigr)
\Bigr)~
\exp\Bigl({\Gamma t \over{ 2}}\bigl({\beta}^2 -
{\beta}^{\dagger 2}\bigr)
\Bigr)~.
\eqno{(2.11)}
$$
\indent In ref. 7 it has been shown that the proper way to perform
the canonical quantization of the damped harmonic oscillator
is to work in the framework
of Quantum Field
Theory .
For many degrees of freedom ${\cal U}(t)$
 is formally (at finite volume) given by
$$
\eqalign{
{\cal U}(t) =
\prod_{\kappa}{\exp\Bigl(-{\Gamma_{\kappa} t \over{ 2}}\bigl(
{\alpha}_{\kappa}^2 -
{\alpha}_{\kappa}^{\dagger 2}\bigr)
\Bigr)
\exp\Bigl({\Gamma_{\kappa} t \over{ 2}}\bigl(
{\beta}_{\kappa}^2 -
{\beta}_{\kappa}^{\dagger 2}\bigr)
\Bigr)}
\cr
=\prod_{\kappa}{\exp{\Bigl ( \Gamma_{\kappa} t \bigl ( A_{\kappa}^{\dagger}
B_{\kappa}^{\dagger} - A_{\kappa} B_{\kappa} \bigr ) \Bigr )}}.}
\eqno{(2.12)}
$$
We still have $[\, H_{0} , H_{I}\, ] = 0$ and the ground state is
(at finite volume $V$)
$$
|0(t)> = \prod_{\kappa} {1\over{\cosh{(\Gamma_{\kappa} t)}}} \exp{
\left ( \tanh {(\Gamma_{\kappa} t)} A_{\kappa}^{\dagger}
B_{\kappa}^{\dagger} \right )} |0> \quad ,
\eqno{(2.13)}
$$
\noindent with
$<0(t) | 0(t)> = 1~, \quad \forall t $~.
Using the continuous limit relation $
\sum_{\kappa} \mapsto {V\over{(2 \pi)^{3}}} \int \! d^{3}{\kappa}$,
in the
infinite-volume limit we have (for $\int \!
d^{3} \kappa ~
\Gamma_{\kappa}$ finite and positive)
$$
{<0(t) | 0> \rightarrow 0~~ {\rm as}~~ V\rightarrow \infty }
{}~~~\forall~  t~  ,
$$
$$
{<0(t) | 0(t') > \rightarrow 0~~ {\rm as}~~ V\rightarrow \infty}
{}~~~\forall~t~ and~ t'~ ,~~~ t' \neq t. \eqno{(2.14)}
$$
At each time $t$ a {\it bona fide} representation
$\{ |0(t)> \}$ of the canonical commutation relations
is defined and turns out to be
unitarily inequivalent  to any other
representation $\{ |0(t')>~,~~\forall t'\neq t \}$ in the infinite volume
limit. In such a way the quantum
damped harmonic oscillator  evolves in time through
unitarily inequivalent
representations of canonical commutation relations ({\it tunneling}).
As usual one works at finite volume and only at the end
of the computations the limit $V \to \infty$ is performed.

Finally, in refs. 6 and 7 it has been shown that the
representation $\{|0(t)>\}$ is equivalent to the
thermo field dynamics  representation
$\{|0(\beta(t)>~,~~\beta(t) \equiv 1/{KT(t)}\}$,
thus recognizing the relation between the damped harmonic oscillator
states and the finite temperature states.

In particular, one may introduce the {\sl free energy}
functional for the $A$-modes
$$
{\cal F}_{A} \equiv <0(t)| \Bigl (  H_{A} - {1\over{\beta}} S_{A}
\Bigr ) |0(t)> \quad , \eqno{(2.15)}
$$
where $H_{A}$ is the part of $H_{0}$ relative to  $A$-
modes only,
namely $H_{A} = \sum_{\kappa} \hbar \Omega_{\kappa}
A_{\kappa}^{\dagger} A_{\kappa}$, and the {\it entropy} $S_{A}$
is given by
$$
 S_{A} \equiv - \sum_{\kappa} \Bigl \{ A_{\kappa}^{\dagger} A_{\kappa}
\ln \sinh^{2} \bigl ( \Gamma_{\kappa} t \bigr ) - A_{\kappa}
A_{\kappa}^{\dagger} \ln \cosh^{2} \bigl ( \Gamma_{\kappa} t \bigr ) \Bigr \}
\quad .  \eqno{(2.16)}
$$
One then considers
the stability condition
${{\partial {\cal F}_{A}}\over{\partial \vartheta_{\kappa}}} = 0 \quad
 \forall \kappa \quad ,\vartheta_{\kappa} \equiv \Gamma_{\kappa} t$~
to be satisfied in each representation,
and using the definition $E_{\kappa} \equiv \hbar \Omega_{\kappa}$, one finds
$$
{\cal N}_{A_{\kappa}}(t) = \sinh^{2} \bigl ( \Gamma_{\kappa} t \bigr ) =
{1\over{{\rm e}^{\beta (t) E_{\kappa}} - 1}} \quad , \eqno{(2.17)}
$$
\noindent which is the Bose distribution for $A_{\kappa}$ at time $t$.
$\{ |0(t)> \}$ is  thus recognized$^{[7]}$ to be a representation of
the canonical commutation relations at finite temperature, equivalent
to the
thermo field dynamics  representation $\{ |0(\beta )> \}$~$^{[11,12]}$.

\bigskip
\bigskip
\noindent {\bf 3. The quantum deformation of the Weyl-Heisenberg algebra}
\bigskip

We now introduce the realization of quantum Weyl-Heisenberg algebra in terms
of finite difference operators$^{[3]}$
in order to establish a formal relation
with the damped harmonic oscillator  hamiltonian.

Since we want to preserve the analytic properties of
Lie algebra in the deformation procedure, we adopt
as a framework the Fock-Bargmann representation
in Quantum Mechanics $^{[15]}$
(we observe that, by working in the Fock-Bargmann representation,
quantum Weyl-Heisenberg algebra is incorporated
into the theory of the (entire) analytic functions, a result
which is by itself interesting$^{[3]}$).

In the Fock-Bargmann representation the operators
$$
N \to z {d\over dz}~,~~
a^\dagger \to z~,~~ a \to {d\over dz}~,~~
z~ \in {\CcC}~~, \eqno{(3.1)}
$$
provide a realization of the Weyl-Heisenberg algebra
$$
[ a, a^\dagger ] = \IiI~,~~ [ N, a ] = - a~,~~ [ N,
a^\dagger ] = a^\dagger~~~. \eqno{(3.2)}
$$

The Hilbert space ~$\cal F$ is identified with the space
of the entire
analytic functions and has well defined inner product. The
wave functions $\psi(z)$ are expressed as~
$$
 \psi (z) = \sum_{n=0}^\infty c_n u_n(z)~,~~
u_n(z) = {z^n\over \sqrt{n!}}~,~~ n~ \in {\cal I}_+~~. \eqno(3.3)
$$
The set ~$\{ u_n(z)\}$~ provides an orthonormal basis in ~$\cal F$.
Eq. (3.3) provides the most general representation of an
entire analytic function. The conjugation of the operators introduced
in eq. (3.1) is defined with respect to the inner product
in $\cal F$ $^{[15]}$.

Next we consider the finite difference operator ${\cal D}_q$
defined by:
$$
{\cal D}_q ~f(z) ~=~ {{f(q z) - f(z)}\over {(q-1) ~z}}~=~
{{q^{z {d\over {dz}}} - 1}\over{(q-1)~ z }}f(z)~, \eqno{(3.4)}
$$
with ~$f \in {\cal F} ,~ q = e^\zeta , ~\zeta \in {\CcC}$.
${\cal D}_q$ is the well known
quantum derivative operator and, for $q \to 1$ (i.e.
$\zeta \to 0$), it reduces to the standard derivative.

We can write the algebra
$$\bigl[ {\cal D}_q~, ~z \bigr] ~=~~ q^{z {d\over {dz}}}~~,
{}~~~~\bigl[ z {d\over dz}, ~{\cal D}_q \bigr] ~=~~ - {\cal D}_q~~,
{}~~~~\bigl[ z {d\over dz}, ~z \bigr] ~=~~ z ~~,
 \eqno{(3.5)}$$
which is recognized to be the quantum deformation of the Weyl-Heisenberg
algebra.
In fact, let us introduce the following operators
in the space ${\cal F}$:
$$
N \to ~z {d\over dz} ~~, ~~~~{\hat a}_q ~\to ~z~~,
{}~~~~~a_q ~\to ~{\cal D}_q ~~, \eqno{(3.6)}
$$
where  ~${\hat a}_q = {\hat a}_{q=1} =
a^\dagger$ and ~$\displaystyle{\lim_{q\to1} a_q =a}$.
The quantum Weyl-Heisenberg algebra, in terms of the operators
$\{ a_q, {\hat a}_q, N\equiv N_q ;
 ~q \in \CcC \}$,  is then realized by the relations:
$$
[ N, a_q ] = - a_q~~,~~[ N, {\hat a}_q ] =
{}~~{\hat a}_q \; ,\;
[ a_q, {\hat a}_q ]~\equiv~a_q {\hat a}_q~-~{\hat a}_q a_q~=
{}~q^N  . \eqno{(3.7)}
$$
By introducing ~~${\bar a}_q ~\equiv ~
{\hat a}_q q^{-N/2}$~, it assumes the more familiar form$^{[2]}$
$$
[ N , a_q ] =~- a_q~~,~~[ N , {\bar a}_q ] =~{\bar a}_q \;
,  \;a_q {\bar a}_q ~- q^{-{1\over 2}} ~{\bar a}_q a_q~=~q^{{1\over
2}N}\; . \eqno{(3.8)}
$$
The finite difference operator algebra (3.7) thus provides
a realization of the quantum Weyl-Heisenberg
algebra in the Fock-Bargmann representation.

We can show that the commutator  $[ a_q, {\hat a}_q]$
acts in ${\cal F}$ as follows$^{[3]}$

$$ [ a_q, {\hat a}_q ]f(z)~=~q^{z{d\over dz}}f(z)~=~f(qz)~.
\eqno{(3.9)}$$

The quantum deformation of
the Weyl-Heisenberg algebra is thus strictly related with
the finite difference operator ~${\cal D}_q$ ($q \not= 1$).
This suggests to us that
the quantum deformation of the operator algebra
should arise whenever we are
in the presence of
lattice or discrete
structure$^{[3]}$.
In the following we show that this indeed happens in the case of damping
where the finite life-time
$\tau = {1 \over{\Gamma}}$
acts as the time-unit for the system.

\bigskip
\bigskip
\noindent {\bf 4. Dissipation, finite temperature
and quantum Weyl-Heisenberg algebra }
\bigskip

In this section we finally express the time evolution
generator of damped harmonic oscillator
and the generator of the finite temperature
Bogolubov transformations in terms  of the commutator
$ [ a_q, {\hat a}_q ]$ of the quantum Weyl-Heisenberg algebra.

In the Fock-Bargmann representation
Hilbert space $~{\cal F}$ we have the identity
$$
2 z {d\over {dz}} f(z) = - \bigl({ \tilde a}^2 -
{ \tilde a}^{\dagger 2}\bigr) f(z) - f(z),
{}~~f \in {\cal F} ~
\eqno{(4.1)}
$$
with
$$
 \tilde a \equiv {1\over {\sqrt{2\hbar\Omega}}} \bigl({ p_z\over{\sqrt{m}}}
- i{\sqrt{m} \Omega} z
\bigr) , ~~ { \tilde a}^\dagger \equiv {1\over {\sqrt{2\hbar\Omega }}}
\bigl({ p_z\over{\sqrt{m}}} +  i{\sqrt{m}\Omega} z
\bigr), ~~~
[\tilde a, {\tilde a}^\dagger ] = \IiI   \eqno{(4.2)}
$$
where $z \in {\CcC}$, $p_z = - i \hbar {d\over {dz}}$ and $[ z, p_z]
= i \hbar$.
The conjugation of $\tilde a$ and ${\tilde a}^\dagger$ is as usual well
defined with respect to the inner product defined in $~{\cal F}$$^{[15]}$.

We note that, by setting $ Re(z)=x$, ${\tilde a} \to \alpha$ and
${\tilde a}^{\dagger} \to {\alpha}^{\dagger}$ in the limit
$Im(z) \to 0$, where $\alpha$ and ${\alpha}^{\dagger}$ denote
the annihilation
and creation operators introduced in eqs. (2.2).

By putting $q={\rm e}^{\theta}$ with $\theta$ real and by using eq. (4.1)
we can easily check that the operator
$$
[ a_q, {\hat a}_q]  ~=~
  \exp\Bigl(\theta z {d\over dz}\Bigr) ~=~
{1\over{\sqrt q}} \exp\Bigl(-{\theta\over{ 2}}\bigl({\tilde a}^2 -
{\tilde a}^{\dagger 2}\bigr)
\Bigr)
\equiv {1\over{\sqrt {q}}}{\hat {\cal S}}(\theta) ,~~
\eqno{(4.3)}
$$
generates the Bogolubov transformations:
$$
\tilde a(\theta) = {\hat{\cal S}}(\theta)
\tilde a {\hat {\cal S}}(\theta)^{-1} =
 \tilde a \cosh{\theta} - \tilde a^{\dagger} \sinh{
\theta}
\eqno{(4.4)}
$$
and h.c..
We then observe that in the limit $Im(z) \to 0$ eq. (4.4) and its h.c. give
$$
\tilde a(\theta) \mapsto \alpha(\theta)  =
 \alpha \cosh{\theta} - {\alpha}^{\dagger} \sinh{
\theta}
{}~~ as~~ Im(z) \to 0
\eqno{(4.5)}
$$
and h.c..

We also observe that
the right hand side of (4.3) is an $SU(1,1)$ group element.
By defining ${1\over  2}{\tilde a} ^2 = K_{-}$,
${1\over  2}{\tilde a}^{\dagger 2} = K_{+}$,
${1\over 2}(\tilde a^\dagger
\tilde a + {1\over 2})
= K_{3}$,
we easily see they close the $su(1,1)$ algebra.
We remark that the transformation (4.5), which is a canonical transformation,
is exactly the one which relates the Weyl representations of the
canonical commutation relations  in
Quantum Mechanics $^{[4]}$
and thus the deformation parameter $q={\rm e}^{\theta}$
acts as a label for the Weyl representations.

Next, we introduce the quantum Weyl-Heisenberg
algebra operators $b_{q'}$ and ${\hat b}_{q'}$
corresponding to the doubled degree of freedom $\beta$ introduced
in Sec. 2 in order to set up the canonical quantization of the
damped harmonic oscillator;
let $\tilde b$ and ${\tilde b}^{\dagger}$ be defined, in a similar way
to  $\tilde a$ and ${\tilde a}^{\dagger}$, by
$$
 \tilde b \equiv {1\over {\sqrt{2\hbar\Omega}}} \bigl({ p_{\zeta}\over{
\sqrt{m}}}
- i{\sqrt{m} \Omega} {\zeta}
\bigr) , ~~ { \tilde b}^\dagger \equiv {1\over {\sqrt{2\hbar\Omega }}}
\bigl({ p_{\zeta}\over{\sqrt{m}}} +  i{\sqrt{m}\Omega} {\zeta}
\bigr), ~~~
[\tilde b, {\tilde b}^\dagger ] = \IiI   \eqno{(4.6)}
$$
with $\zeta \in {\CcC}$, $Re(\zeta) \equiv y$,
$p_{\zeta} = - i \hbar {d\over {d \zeta}}$ and
$[ \zeta, p_{\zeta}] = i \hbar$, so that
$$
2 \zeta {d\over {d \zeta}} f(\zeta) = - \bigl({ \tilde b}^2 -
{ \tilde b}^{\dagger 2}\bigr) f(\zeta) - f(\zeta),
{}~~f \in {\cal F} ~
\eqno{(4.7)}
$$
and ${\tilde b} \to {\beta}$, ${\tilde b}^{\dagger} \to {\beta}^{\dagger}$
as $Im(\zeta) \to 0$.
Then we see that the operator
$$
[ a_{q}, {\hat a}_{q}][ b_{q'}, {\hat b}_{q'}] =
\exp \Bigl(-{{\theta} \over 2}
\bigl[\bigl({{\tilde a}} ^2 -
{{\tilde a}}^{\dagger 2} \bigr) - \bigl({{\tilde b}}^2 -
{{\tilde b}}^{\dagger 2} \bigr)\bigr] \Bigr),
\eqno{(4.8)}
$$
with $q'={q}^{-1}$, acts as the time evolution operator
${\cal U}(t)$
(cf. eq. (2.12)) in the limits $Im(z) \to 0$ and $Im(\zeta) \to 0$
provided
we set $q={\rm e}^{\theta}$, with ${\theta} \equiv {\Gamma} t$:
$$
\eqalign{
[ a_{q}, {\hat a}_{q}][ b_{q'}, {\hat b}_{q'}]
\mapsto
\exp \Bigl(-{{\Gamma t} \over 2}
\bigl[\bigl({{\alpha}} ^2 -
{{\alpha}}^{\dagger 2} \bigr) - \bigl({{\beta}}^2 -
{{\beta}}^{\dagger 2} \bigr)\bigr] \Bigr)=
{\cal U}(t)  \cr
as~~Im(z) \to 0~~and~~Im(\zeta) \to 0~~~.}
\eqno{(4.9)}
$$
\indent Eq. (4.9) is the wanted expression of
 the time-evolution generator of the damped harmonic oscillator
in terms of the quantum Weyl-Heisenberg
algebra operator $[ a_{q}, {\hat a}_{q}][ b_{q'}, {\hat b}_{q'}]$.

For many degrees of freedom, the Bogolubov transformations,
corresponding to eqs. (2.10),
can be implemented for every $\kappa$ as inner automorphism for the
algebra  ${su(1,1)}_{\kappa}$. As shown in ref. 7, at every time $t$
we have a copy
$\{ A_{\kappa}(t) , A_{\kappa}^{\dagger}(t) , B_{\kappa}(t) ,
B_{\kappa}^{\dagger}(t) \, ; \, | 0(t) >\, |\, \forall {\kappa} \}$
of the
original algebra induced by the time evolution operator which
can thus be thought of as a generator
of the group of automorphisms of $\displaystyle{\bigoplus_{\kappa}
su(1,1)_{\kappa}}$ parameterized by time $t$ (we have a
realization of the operator algebra at
each time t,
which can be implemented by Gel'fand-Naimark-Segal construction in the
C*-algebra formalism).
Notice that the various copies
become unitarily inequivalent in the infinite-volume limit, as
shown by eqs. (2.14): the space of the states splits
into unitarily inequivalent
representations of the
canonical commutation relations  each one labelled by time
parameter $t$.
Our discussion can be then generalized to the case of many degrees of
freedom by observing that eq. (4.9)
holds for each couple of modes (${\alpha}_{\kappa},{\beta}_{\kappa}$)
and formally we have
$$
\prod_{\kappa}{[ a_{\kappa,q}, {\hat a}_{\kappa,q}][ b_{\kappa,q'},
{\hat b}_{\kappa,q'}]} ~\to~
\exp\bigl(-{i\over{\hbar}} H_{I}t \bigr)={\cal U}(t)~~, ~~~
as~Im\{z,\zeta \}  \to 0~~,~
\eqno{(4.10)}
$$
where $q={\rm e}^{\theta_\kappa}$,
${\theta_\kappa} \equiv {\Gamma_\kappa}t$,  $q'={q}^{-1}$,
$Im\{z,\zeta\}$ denotes the imaginary part of Fock-Bargmann representation
$z$- and
$\zeta$-variables associated
to each ${\alpha}_{\kappa}$ and ${\beta}_{\kappa}$ mode and ${\cal U}(t)$
is given by eq. (2.12).
Notice that for simplicity of notation here $q$ denotes $q_{\kappa}$.

We can then conclude that, through its time
dependence, the deformation
parameter $q(t)$ labels the
unitarily inequivalent  representations $\{|O(t)>\}$.

On the other hand, we also observe that eq. (4.10) leads to
representation of the generator
of the thermal Bogolubov transformation in terms of the operators
$\prod_{\kappa}{[ a_{\kappa,q}, {\hat a}_{\kappa,q}]
[ b_{\kappa,q'}, {\hat b}_{\kappa,q'}]}$:
indeed, by resorting to the discussion of Sec. 2 where the representation
$\{|O(t)>\}$ for the damped harmonic oscillator has been
shown to be the same as the
thermo field dynamics  representation $\{|O(\beta (t))>\}$ (see refs. 6-7),
we also recognize the strict relation between
quantum Weyl-Heisenberg algebra and
finite temperature Quantum Field
Theory .

As a matter of fact, from eq. (2.17) we
can set ${\theta_\kappa}(t) \equiv {\theta_\kappa}(\beta (t))$ and
the operator ${\cal G} \equiv \prod_{\kappa}
\exp\left[ {\theta}_{\kappa} (\beta (t))
\bigl  ({{A}_{\kappa}} ^{\dagger}
{{B}_{\kappa}} ^{\dagger} -
{A}_{\kappa}{B}_{\kappa} \bigr )\right]$ indeed implements the
(time-dependent) finite temperature Bogolubov transformation
in thermo field dynamics $^{[7,11,12]}$.

We finally remark that the derivation of
eq. (4.10) (and (4.8)) holds even independently of
the above discussion of the damped harmonic oscillator  quantization
and, by setting $\theta_{\kappa} =
\theta_{\kappa}(\beta)$, we see that the operator
$ \prod_{\kappa} [ a_{\kappa,q}, {\hat a}_{\kappa,q}][ b_{\kappa,q'},
{\hat b}_{\kappa,q'}]$
leads to the generator of the Bogolubov transformations in
conventional thermo field dynamics  in the $Im\{z,\zeta\} \to 0$ limit:
quantum Weyl-Heisenberg algebra thus appears as the natural
candidate to study thermal field theories, too.

\bigskip
\bigskip
\noindent {\bf 5. Conclusions}
\bigskip

We have shown that
$
\prod_{\kappa}{[ a_{\kappa,q}, {\hat a}_{\kappa,q}][ b_{\kappa,q'},
{\hat b}_{\kappa,q'}]}
$
acts as time-evolution operator
for the damped harmonic oscillator
and as generator of thermal Bogolubov transformation
in the $ Im\{z,\zeta\} \to 0$ limit.

In the discussion presented above a crucial r\^ole is
played by the existence of infinitely many
unitarily inequivalent  representations of the
canonical commutation relations
in Quantum Field
Theory . In ref. 4 the quantum Weyl-Heisenberg algebra has been discussed in
relation with the von Neumann theorem in Quantum Mechanics  and it has been
shown on a general ground that
the quantum deformation parameter acts as a label for the
{\it Weyl systems} in Quantum Mechanics  and for the
unitarily inequivalent  representations in Quantum Field
Theory; the
mapping between {\it different}
 ($i.e.$ labelled by different values of $q$)
representations (or Weyl systems) being
performed by the Bogolubov transformations (at finite volume).
In this paper we have shown by the explicit examples of damped
harmonic oscillator and
finite temperature systems the physical meaning of such a labelling
(further examples are provided by unstable particles in Quantum Field
Theory $^{[17]}$,
by
quantization of the matter field in curved space-time$^{[18]}$, by
theories with spontaneous breakdown of symmetry where  different
values of the order parameter are associated to
different
unitarily inequivalent representations (different {\it phases})).
In the case of damping,
as well as in the
case of time-dependent temperature,
the system time-evolution is represented as
{\it tunneling} through unitarily inequivalent
representations: the non-unitary
character of time-evolution ({\it arrow of time}) is thus expressed
by the non-unitary equivalence of the representations in the infinite
volume limit. It is remarkable that at the algebraic level this is
made possible through the
quantum deformation mechanism which
organizes the representations in an {\it ordered set} by means of
the labelling.

In ref. 3 it has been also shown that
the commutator $[ a_q, {\hat a}_q]$ acts as squeezing
generator (indeed the operator ${\hat S}(\theta)$ in eq. (4.3) acts like
the squeezing generator with respect to ${\tilde a}$ and ${\tilde a}^\dagger$
operators), a result
which together with the conclusions of this paper confirms
the relation between dissipation and squeezed coherent states
exhibited in ref. 5. In turn, quantum groups have been also shown$^{[19]}$
to be
the  natural candidates to study squeezed coherent states.

The above remarks strongly suggest to us that quantum deformations
are mathematical structures which are characteristic of basic, deep
features of Quantum Field
Theory  and much work still needs to be done to fully
investigate the relation between Quantum Field
Theory  and quantum groups. For example,
in this paper we have not studied the r\^ole played
by the coproduct operation of the quantum deformation. We plan to
study this subject in the future.

We are glad to acknowledge useful and enjoying
discussions with E.Celeghini, S.De
Martino, S.De Siena, V. Man'ko and M.Rasetti.

\vfill
\eject

\bigskip
\bigskip

\phantom{xxxxx}
\bigskip
{\bf References}

\bigskip

\baselineskip= 16 pt

\ii {1} Drinfeld V.G., Proc. ICM Berkeley, CA; A.M. Gleason, ed,; AMS,
	   Providence, R.I., 1986, page 798.

\ii	{\phantom{1}} Jimbo M., Int. J. of Mod. Phys. {\bf A4} (1989) 3759.

\ii	{\phantom{1}} Manin Yu.I., {\it
        Quantum groups and Non-Commutative Geometry},
	   Centre de Recherches Math\'ematiques, Montreal, 1988.

\ii 2   L.C.Biedenharn, {\it J.Phys.}  {\bf A22} (1989) L873

\ii	{\phantom{2}} A.J.Macfarlane, {\it J. Phys.} {\bf A22}
      	(1989) 4581

\ii 3   E.Celeghini, S.De Martino, S.De Siena, M.Rasetti and
        G. Vitiello, {\it Mod. Phys. Lett.} {\bf B 7} (1993) 1321;
        {\it Quantum Groups, Coherent States, Squeezing
	and Lattice Quantum Mechanics}, preprint 1993

\ii 4   A.Iorio and G.Vitiello, {\it Mod. Phys. Lett.} {\bf B 8} (1994) 269

\ii 5	E. Celeghini, M. Rasetti, M. Tarlini and
	G. Vitiello, {\sl Mod.
	Phys. Lett.} {\bf B 3} (1989) 1213

\ii 6	E.Celeghini, M.Rasetti and G.Vitiello in {\it Thermal Field
	Theories and Their Applications}, H.Ezawa, T.Aritmitsu
	and Y.Hashimoto Eds., Elsevier, Amsterdam 1991, p. 189

\ii 7	E. Celeghini, M. Rasetti and G. Vitiello, {\it Annals of Phys.
	(N.Y.)} {\bf 215} (1992) 156

\ii 8	Y.N.Srivastava, G.Vitiello and A.Widom, {\it Quantum Dissipation
	and quantum noise}, {\it Annals of Phys. (N.Y.)}, in press

\ii 9   H.P.Yuen, {\it Phys. Rev.} {\bf A13} (1976) 2226

\item{10} 	E.Celeghini, T.D.Palev and M.Tarlini, {\it Mod. Phys.Lett.}
        {\bf B5 } (1991) 187

\item 	{\phantom{10}} P.P.Kulish and N.Yu.Reshetikin, {\it Lett. Math.
	Phys.} {\bf 18} (1989) 143

\item{11}  Y. Takahashi and H. Umezawa, {\sl Collective Phenomena} {\bf 2}
	(1975) 55

\item{12} H. Umezawa, H. Matsumoto and M. Tachiki, {\it
	Thermo Field Dynamics
        and Condensed States}, North-Holland Publ.
        Co., Amsterdam 1982

\item{13} A.Iorio and G.Vitiello,
        in {\it Banff/CAP Workshop on Thermal
        Field Theory}, F.C. Khanna et al. eds., World Sci., Singapore
        1994, p.71

\item{14} H.Feshbach and Y.Tikochinsky, {\sl Transact. N.Y. Acad. Sci.}
	{\bf 38} (Ser. II) (1977) 44

\item{15} A.Perelomov, {\it Generalized Coherent States and Their
        Applications}, Springer-Verlag, Berlin, Heidelberg 1986

\item{16} J.R.Klauder and E.C.Sudarshan, {\it Fundamentals of Quantum
	Optics}, Benjamin, New York 1968

\item{17} S. De Filippo and G. Vitiello,
{\sl Lett. Nuovo Cimento} {\bf 19} (1977) 92

\item{18} M. Martellini, P. Sodano and G. Vitiello,
{\sl Nuovo Cimento} {\bf 48 A} (1978) 341

\item{19} E.Celeghini, M.Rasetti and G.Vitiello, {\it Phys. Rev.
	Lett.} {\bf 66} (1991), 2056

\vfill\eject
\bye